\documentclass[conference]{IEEEtran}
\IEEEoverridecommandlockouts
\usepackage{cite}
\usepackage{amsmath,amssymb,amsfonts}
\usepackage{algorithmic}
\usepackage{graphicx}
\usepackage{textcomp}
\usepackage{xcolor}
\def\BibTeX{{\rm B\kern-.05em{\sc i\kern-.025em b}\kern-.08em
    T\kern-.1667em\lower.7ex\hbox{E}\kern-.125emX}}
\begin{document}

\title{Covert Transmission with Antenna Selection and Using an External Jammer\\}

\author{\IEEEauthorblockN{ Morteza Sarkheil}
\IEEEauthorblockA{\textit{Department of ECE,} \\ \textit{Tarbiat Modares University} \\
Tehran, Iran \\
morteza.sarkheil@modares.ac.ir}
\and
\IEEEauthorblockN{ Paeiz Azmi}
\IEEEauthorblockA{\textit{Department of ECE,} \\ \textit{Tarbiat Modares University} \\
Tehran, Iran \\
pazmi@modares.ac.ir}
\and
\IEEEauthorblockN{ Moslem Forouzesh}
\IEEEauthorblockA{\textit{Department of ECE,} \\ \textit{Tarbiat Modares University} \\
Tehran, Iran \\
m.Forouzesh@modares.ac.ir}
\and
\IEEEauthorblockN{ Ali Kuhestani}
\IEEEauthorblockA{\textit{Department of ECE,} \\ \textit{Sharif University of Technology} \\
Tehran, Iran \\
kuhestani@sharif.edu}
}
\maketitle
\begin{abstract}
This paper adopts the antenna selection technique to enhance the covert rate in a wireless communication network comprised of
a source, a destination , an external jammer and  an eavesdropper. In the covert communication, the level of transmit power is low and hence a source with multiple antennas can be adopted to send the information toward the single antenna destination while concurrently, the jammer transmits an artificial noise signal. For this system model, we consider a scenario where the source is forced to select one or several of its antennas to transmit its confidential information due to its limited RF chains. Furthermore, we consider two different jamming scenarios to support our covert communication: 1) The destination is unable to cancel the jamming signal, 2) The destination can subtract the jamming signal. For such a communication network, our aim is to maximize the covert rate subject to power constraint and covert communication requirement. In the first scenario, the optimization problem is non-convex, and hence, it can be solved through using Difference of Convex function (DC) method while the optimization problem of the second scenario is intrinsically convex. 
Our numerical results show that the higher the number of selected antennas at  the transmitter, the higher the covert rate will be achieved.
\end{abstract}


\section{Introduction}
Security is a critical and important subject in wireless communications networks. This is because the broadcast nature of wireless networks permits to the unauthorized nodes to access the contents of the confidential messages [1]. 
Physical layer security as a new solution to enhance the confidentiality of wireless communications has attracted a lot of interest [2]--[4]. Physical layer secure transmission is provisioned by intelligently exploiting the time varying properties of fading channels, instead of relying on conventional cryptographic techniques. This approach uses signal processing and encoding techniques at the physical layer, to improve the quality of the received signal at illegitimate receivers compared with the unauthorized users. Toward this end, the jamming signal can be used to enhance the physical layer security. Typically, there are two main types of jamming signal to enhance the security of wireless networks [4], [5]: 1) Friendly jamming (FJ) scenario where the jamming signal is known at the legitimate receiver, 2) Gaussian noise jamming (GNJ) where the jamming signal is unknown at the legitimate receiver[5]-[7]. It should be noted that FJ provides better secrecy performance compared with GNJ because when FJ is used, legitimate  receiver can cancel the jamming signal. However, GNJ is more simple compared with FJ scenario. This reveals the trade-off between secrecy performance and complexity of the network.

In some communication networks, low probability of detection(LPD) or covert communication is essential  for information transmission over electromagnetic and acoustic channels [8]. For military applications, LPD is interest when the transmitter wishes to remain undetected, or when the knowledge of communication may point to the presence of a receiver. As such, in covert communication only the detection capability of an eavesdropper is considered, i.e., an eavesdropper need not be able to actually decode the communication signal. In other words, the covert communication keeps military forces from possible attacks [9]. In recent years, some few works have investigated the covert communication in different wireless communication networks [9]--[14]. 

In this paper, we take into account the power allocation problem of a wireless communication network, where a multiple antenna source transmits its confidential message to a single antenna destination in the presence of a passive eavesdropper. To covert the communication against the eavesdropping attack, a single antenna external jammer is employed. For this communication network, we investigate two different jamming scenarios to support covert communication: 1) The destination is unable to cancel the jamming signal, i.e., we have FJ, 2) The destination can subtract the jamming signal, i.e., we have GNJ. We consider a realistic scenario where the number of RF chains at the source may be fewer than the number of source's antennas. Instead of selecting the antennas randomly, we propose to select the best antennas to transmit the information signal toward the destination. 
For such a network, we formulate the power allocation between the source
and jammer that maximizes the instantaneous covert
rate while concurrently hiding the
communication against the passive eavesdropping attack.
Since the optimization problem of GNJ scenario is non-convex, we exploit difference of concave (DC) approach to convert it to a convex optimization problem. For the FJ scenario, we observe that the optimization problem is convex. We also obtain closed-form solutions for the optimal power threshold from the eavesdropper's viewpoint. Our numerical examples show that by increasing the number of selected antennas at the transmitter  the covert rate is increased. Furthermore, the impact of the distance between the transmitter and the eavesdropper on the achievable covert rate is more than the impact of the distance between the transmitter and the receiver.

\section{System Model}
As shown in Fig. 1, the system model under consideration
is a multi-input single-output (MISO) wireless network consisting of an $M_T$ antenna transmitter (Alice), a single antenna jammer, a single antenna receiver (Bob), and a single antenna eavesdropper (Eve).
For this system model, assume  Alice selects several antennas to send  information signal and also, the jammer sends jamming signal in all directions. 
In this scenario, we assume the communication channels follow the slow fading. Moreover, the fading coefficients have Rayleigh distribution. Alice is unaware of Eve's channel state information (CSI) and only knows the distance between itself and Eve. We also assume that the jammer knows the distance between itself and Eve. The distance between Alice and Bob, Alice and Eve, the jammer and Bob and the jammer and Eve are defined as ${D_{ab}}$, ${D_{ae}}$, ${D_{jb}}$ and ${D_{je}}$, respectively. The information signal and the jamming signal denoted by ${\boldsymbol{X_s}}=\left[ {{x_s}^1,{x_s}^2,..,{x_s}^m} \right]$ and ${\boldsymbol{X_j}}=\left[ {{x_j}^1,{x_j}^2,..,{x_j}^m} \right]$, respectively, where $m$ shows the number of time slots.
\begin{figure}[t!]
	\begin{center}
		\centering
		\includegraphics[width=3.5in]{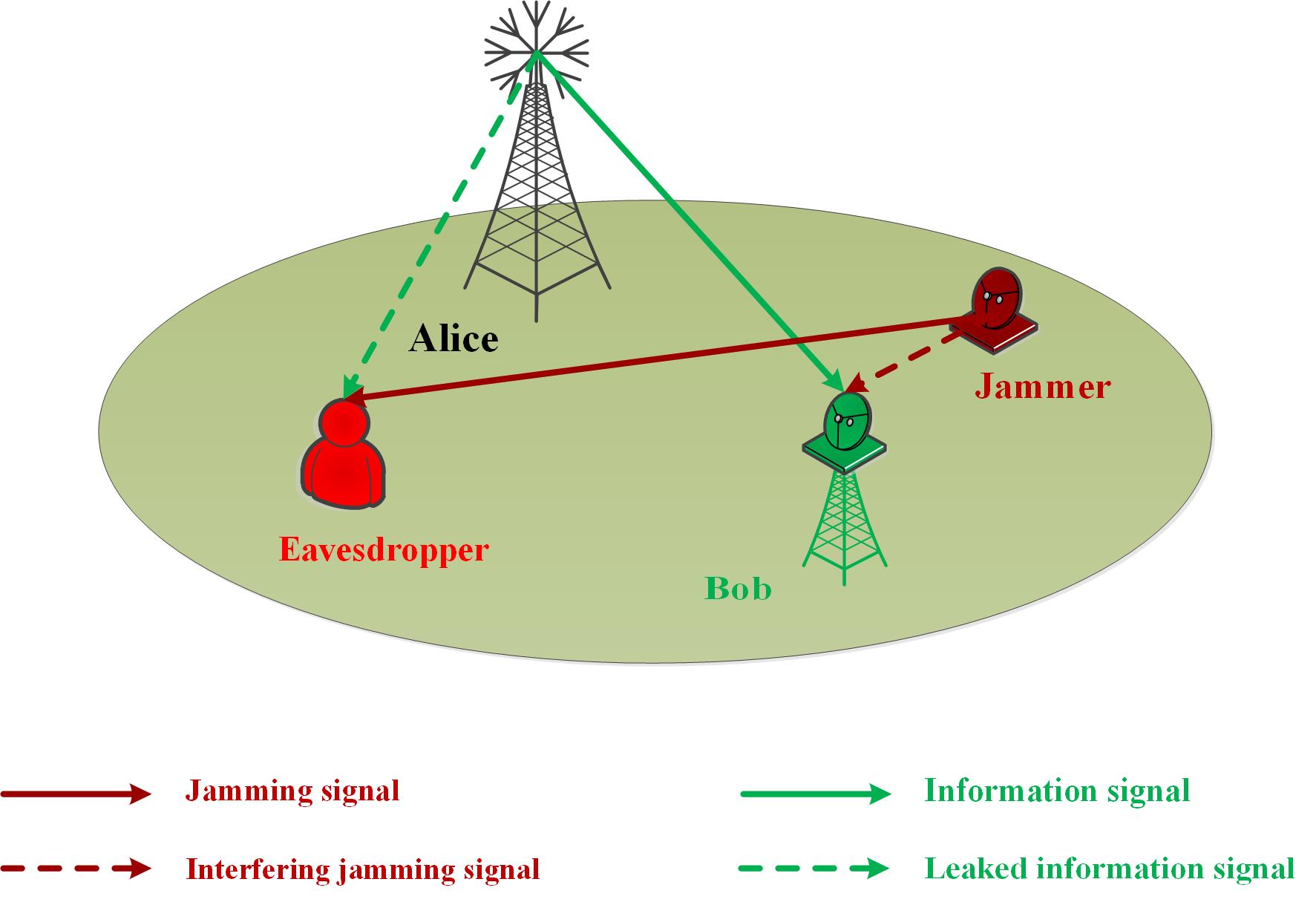}
		\caption{Our MISO system model in the presence of an external jammer and a passive eavesdropper.}
		\label{k}
	\end{center}
\end{figure}
\section{Covert Communication}
In the covert communication, our aim is to send a message from Alice to Bob secretly such that Eve will not be notified of this communication, i.e., based on the level received energy. The Eve decides about the presence or absence of signal transmission by Alice.
In this scenario, notation ${\Omega _0}$ states that Alice does not transmit the information signal to Bob, while ${\Omega _1}$ states that the Alice transmits the information signal to Bob. If we assume $P({\Omega _0}) = 1 - p$ and $P({\Omega _1}) = p$, the probability of error at Eve according to its decisions is defined as follows
\begin{eqnarray}\label{e1}
&&{P_e} = (1 - p){P_{FA}} + p{P_{MD}},
\end{eqnarray}
where ${P_{MD}}$ is the probability that Alice sends its information signal but Eve decides on the absence of communication. Furthermore,  ${P_{FA}}$ is the probability that Alice is silent but Eve decides on the presence of communication between Alice and Bob. The covert communication between Alice and Bob is established when the following condition holds [12]
\begin{eqnarray}\label{e2}
&&\mathop {{P_{FA}} + {P_{MD}}}\limits_{n \to \infty }  \ge 1 - \varepsilon. 
\end{eqnarray}
The received signal at node $k\in{Bob, Eve},$ is as follows

\begin{eqnarray}\label{e3}
{y_k} = \left\{ {\begin{array}{*{20}{l}}
	{\frac{{\sqrt {{p_j}} {h_{jk}}{x_j}}}{{D_{jk}^{\beta /2}}} + {n_k},\begin{array}{*{20}{c}}
		{}&{}&{\,\,\,\,\,\,\,\,\,\,\,\,\,\,\,\,\,\,\,\,\,\,\,\,\,\,}&{{\Omega _0,}}&
		\end{array}}\\
	{\frac{{\sqrt {{p_j}} {h_{jk}}{x_j}}}{{D_{jk}^{\beta /2}}} + \frac{{\sqrt {{p_s}} {\boldsymbol{w}^H}\bf{{h_{ak}}}{x_s}}}{{D_{ak}^{\beta /2}}} + {n_k},\begin{array}{*{20}{c}}
		{}&{{\Omega _1},}
		\end{array}}
	\end{array}} \right.
\end{eqnarray}
where $p_s$ and $p_j$ are the allocated power for the information signal and the jamming signal, respectively, such that ${p_s} = \alpha {P_{total }}$ and ${p_j} = (1 - \alpha ){P_{total }}$, where $\alpha$ is the power allocation factor. It is worth noting that the total power consumption constraint
provides a guideline for the power allocation of the Alice 
and the jammer. This approach has been widely exploited in
the literature for both performance analysis and network
optimization design [7], [11], [14]. Furthermore, in (3), $\beta$  is the path-loss exponent. Note that $\boldsymbol{w} = \frac{{{\bf{h_{ab}}}}}{{\left\| {\left. {\bf{{h_{ab}}}} \right\|} \right.}}$represents the maximum ratio transmission
(MRT) beamformer at the source and $n_k$  is the white Gaussian noise received at node $k$ with the distribution of $n \sim CN(0,\sigma _k^2)$. ${\bf{{h_{ak}}}}$, and ${{\bf{h_{jk}}}}$ are the channel coefficients  of the information signal and the jamming signal, respectively.

The decision of Eve is based on the energy of the received signal. Eve decides on the state $\Omega_1$ if $\frac{{{Y_e}}}{m} \ge V$. Otherwise, Eve decides on the state $\Omega_0$.
$V$ is the threshold of decision and $Y_e$ is the energy of the received signal which is ${Y_e} = {\sum\limits_{l = 1}^m {\left| {y_k^l} \right|} ^2}$. Accordingly, $P_{FA}$ and $P_{MD}$ can be written, respectively, as
\begin{eqnarray}\label{e6}
\begin{array}{l}
{P_{FA}} = P(({\sigma _e}^2 + \gamma )\frac{{{\chi ^2}_{2m}}}{m} \ge V\left| {{\Omega _0}} \right.),
\end{array}
\end{eqnarray}
\begin{eqnarray}\label{d1}
\begin{array}{l}
{P_{MD}} = P(({\sigma _e}^2 + \gamma )\frac{{{\chi ^2}_{2m}}}{m} \le V\left| {{\Omega _1}} \right.),
\end{array}
\end{eqnarray}
where $\chi^2_{2m}$ is a random variable with chi-squared distribution
with $2m$ degrees of freedom. For large number of time slots, $m \to \infty $, we have $\frac{{\chi _{2m}^2}}{m} \to 1$. Therefore, (5) and (6) can be simplified as 
\begin{eqnarray}\label{d3}
{P_{FA}} = P(({\sigma _e}^2 + \gamma ) \ge V\left| {{\Omega _0}} \right.),
\end{eqnarray}
\begin{eqnarray}\label{d4}
{P_{MD}} = P(({\sigma _e}^2 + \gamma ) \le V\left| {{\Omega _1}} \right.).
\end{eqnarray}
We note that the received signal has the distribution of $y_k^l \sim CN(0,({\sigma ^2} + \gamma ))$, where $\gamma$ is defined as follows

\begin{eqnarray}\label{d5}
\gamma  = \left\{ {\begin{array}{*{20}{l}}
	{\frac{{{P_j}}}{{D_{ae}^\beta }}{{\left| {{h_{je}}} \right|}^2},\begin{array}{*{20}{c}}
		{}&{\,\,\,\,\,\,\,\,\,\,\,\,\,\,\,\,\,\,\,\,\,\,}&{{\kern 1pt} {\kern 1pt} {\kern 1pt} {\kern 1pt} {\kern 1pt} {\kern 1pt} {\kern 1pt} {\kern 1pt} {\kern 1pt} {\kern 1pt} {\kern 1pt} {\kern 1pt} {\kern 1pt} {\kern 1pt} {\kern 1pt} {\kern 1pt} {\kern 1pt} {\kern 1pt} {\kern 1pt} {\kern 1pt} {\kern 1pt} {\kern 1pt} {\kern 1pt} {\kern 1pt} {\kern 1pt} {\kern 1pt} }&{{\Omega _0},}
		\end{array}}\\
	{\frac{{{P_j}}}{{D_{je}^\beta }}{{\left| {{h_{je}}} \right|}^2} + \frac{{{P_s}}}{{D_{ae}^\beta }}{{\left\| {{\boldsymbol{w}^H}\bf{{h_{ae}}}} \right\|}^2},\begin{array}{*{20}{c}}
		{}&{}&{{\Omega _1}.}
		\end{array}}
	\end{array}} \right.
\end{eqnarray}
In \eqref{d5}, $h_{je}$ and $\bf{h_{ae}}$  are zero-mean Gaussian random variables with unit variances. Thus, ${\left| {{h_{je}}} \right|^2}$ has an exponential distribution with the parameter  $\lambda  = 1$. Therefore, for the assumption of ${\Omega _0}$
, $\gamma$ has an exponential distribution with the parameter of $\lambda  =   \frac{1}{{{\phi _0}}}$. According to [15], the summation of two exponential distributions $x$ and $y$ with parameters $\lambda_1$ and $\lambda_2$ is given by
\begin{eqnarray}
{f_{x + y}}(\gamma ) = \frac{{{\lambda _1}{\lambda _2}}}{{{\lambda _2} - {\lambda _1}}}[\,{e^{ - {\lambda _1}\gamma }} - \,{e^{ - {\lambda _2}\gamma }}]U(\gamma),
\end{eqnarray}
where $U(\gamma)$ is the step function. Furthermore, $\gamma$ is distributed as follows
\begin{eqnarray}\label{d6}
{f}(\gamma ) = \left\{ {\begin{array}{*{20}{l}}
	{\begin{array}{*{20}{c}}
		{\frac{1}{{{\phi _0}}}{e^{ - \frac{\gamma }{{{\phi _0}}}}},}&{}&{}&{\begin{array}{*{20}{c}}
			{\,\,\,\,\,\,\,\,\,\,\,\,\,}&{\gamma  \ge 0}&{{\Omega _0,}}
			\end{array}}
		\end{array}}\\
	{\frac{1}{{{\phi _0} - {\phi _1}}}({e^{ - \frac{\gamma }{{{\phi _0}}}}} - {e^{ - \frac{\gamma }{{{\phi _1}}}}}),\begin{array}{*{20}{c}}
		{}&{\gamma  \ge 0}&{{\Omega _1,}}
		\end{array}}
	\end{array}} \right.
\end{eqnarray}
where ${\phi _0} = \frac{{{P_j}}}{{D_{je}^\beta }}$ and ${\phi _1} = \frac{{{P_s}}}{{D_{ae}^\beta }}$. By combining \eqref{d3}, \eqref{d4}, \eqref{d5} and \eqref{d6}, we obtain 
\begin{eqnarray}\label{b1}
{P_{FA}} = \left\{ {\begin{array}{*{20}{l}}
	{{e^{ - \frac{{(V - \sigma _e^2)}}{{{\phi _0}}}}},\begin{array}{*{20}{c}}
		{}&{}&{}&{V - \sigma _e^2 \ge 0,}
		\end{array}}\\
	{1,\begin{array}{*{20}{c}}
		{}&{}&{{\mkern 1mu} {\mkern 1mu} {\mkern 1mu} {\mkern 1mu} {\mkern 1mu} {\mkern 1mu} {\mkern 1mu} {\mkern 1mu} {\mkern 1mu} {\mkern 1mu} {\mkern 1mu} {\mkern 1mu} {\mkern 1mu} {\mkern 1mu} {\mkern 1mu} {\mkern 1mu} {\mkern 1mu} \,\,\,\,\,\,\,\,\,\,\,\,\,\,{\mkern 1mu} {\mkern 1mu} {\mkern 1mu} }&{V - \sigma _e^2 \le 0,}
		\end{array}}
	\end{array}} \right.
\end{eqnarray}
and
\begin{eqnarray}\label{b2}
{P_{MD}} = \left\{ {\begin{array}{*{20}{l}}
	{\Gamma ,\begin{array}{*{20}{c}}
		{}&{\begin{array}{*{20}{c}}
			{V - \sigma _e^2}
			\end{array} \ge 0,}
		\end{array}}\\
	{0,\begin{array}{*{20}{c}}
		{}&{\begin{array}{*{20}{c}}
			{V - \sigma _e^2}
			\end{array} \le 0,}
		\end{array}}
	\end{array}} \right.
\end{eqnarray}
where ${\Gamma} = \frac{1}{{{\phi _0} - {\phi _1}}}({\phi _1}{e^{ - \frac{{(V - \sigma _e^2)}}{{{\phi _1}}}}} - {\phi _0}{e^{ - \frac{{(V - \sigma _e^2)}}{{{\phi _0}}}}} + {\phi _0} - {\phi _1})$.

\subsection{ Eavesdropper's error}
The Eve's error is defined as ${P_{MD}} + {P_{FA}}$. In this scenario, we consider the worst case scenario where Eve has an optimal threshold. The optimal threshold of Eve is minimum decision error. Therefore, we can write
\begin{eqnarray}\label{b3}
{P_{MD}} + {P_{FA}} = \left\{ {\begin{array}{*{20}{l}}
	{{e^{ - \frac{{(V - \sigma _e^2)}}{{{\phi _0}}}}} + {\Lambda _1},\begin{array}{*{20}{c}}
		{}&{V - \sigma _e^2 \ge 0,}
		\end{array}}\\
	{1,\begin{array}{*{20}{c}}
		{\begin{array}{*{20}{c}}
			{\begin{array}{*{20}{c}}
				{}&{}
				\end{array}}&{}&{\,\,\,\,}
			\end{array}}&{V - \sigma _e^2 \le 0,}
		\end{array}}
	\end{array}} \right.
\end{eqnarray}
where 
\begin{eqnarray}\label{b4}
{\Lambda _1} = \frac{1}{{{\phi _0} - {\phi _1}}}({\phi _1}{e^{ - \frac{{(V - \sigma _e^2)}}{{{\phi _1}}}}} - {\phi _0}{e^{ - \frac{{(V - \sigma _e^2)}}{{{\phi _0}}}}} + {\phi _0} - {\phi _1}).
\end{eqnarray}
The minimum Eve's error will be 
\begin{eqnarray}\label{b5}
\mathop {\min }\limits_V ({P_{FA}} + {P_{MD}}) = {e^{ - \frac{{{\phi _1A}}}{{{\phi _1} - {\phi _0}}}}} + {\Lambda _2},
\end{eqnarray}
where 
\begin{eqnarray}\label{b6}
{\Lambda _{2 = }}\frac{1}{{{\phi _0} - {\phi _1}}}({\phi _1}{e^{ - \frac{{{\phi _0}}}{{{\phi _1} - {\phi _0}}}A}} - {\phi _0}{e^{ - \frac{{{\phi _1}}}{{{\phi _1} - {\phi _0}}}A}} + {\phi _0} - {\phi _1}),
\end{eqnarray}
where $A = Ln\frac{{{\phi _1}}}{{{\phi _0}}}$ and the optimal threshold is given by
\begin{eqnarray}\label{q1}
V^* = \frac{{{\phi _0}{\phi _1}}}{{{\phi _1} - {\phi _0}}}Ln\frac{{{\phi _1}}}{{{\phi _0}}} + \sigma _e^2.
\end{eqnarray}
\textit{Proof}: See Appendix A. \hspace{5cm} $\blacksquare$
\subsection{Covert rate}
The covert rate at Bob is defined as 
\begin{eqnarray}\label{q2}
R = \log (1 + \frac{{\alpha {P_{total}}{{\left| {{h_{ab}}} \right|}^2}D_{jb}^\beta }}{{D_{jb}^\beta D_{ab}^\beta \sigma _b^2 + (1 - \alpha ){P_{total}}{{\left| {{h_{jb}}} \right|}^2}D_{ab}^\beta }}).
\end{eqnarray}
Our goal is to increase the covert rate. To this end, from the $M_T$ antennas, $N_D$ of  antennas with the highest channel coefficients between Alice and Bob are selected to send the information. As such, the corresponding antennas are selected based on the following criterion 
\begin{align}\label{e4}
{\bf{h_{ab}}} = {{\rm X}_{{N_D}}}(sort({H_{{M_T} \times 1}})),
\end{align}
where $sort(v)$ arranges the $v$ elements in descending order and ${{\rm X}_i}(u)$ selects first $i$ elements of $u$.
According to \eqref{e4}, we can increase the covert rate by selecting appropriate antennas for sending information signal. The proposed antenna selection technique, not only increases  the covert rate, but also reduces the hardware equipments at Alice, including  amplifiers and mixers, and also reduces the network complexity and overhead.

\section{Optimization Problem}
In this paper, our goal is to  maximize the covert rate subject to the limitation of transmit power and the covert communication condition.  Moreover, in the following, we investigate two different scenarios of FJ and GNJ.
\begin{subsection}{Gaussian Noise Jamming Scenario}
	In  this scenario, it is assumed that Bob cannot cancel the jamming signal. Therefore, we have the following optimization problem
	\begin{eqnarray}\label{q3}
	\begin{array}{*{20}{l}}
	{\mathop {\max }\limits_\alpha  \{ R = \log (1 + \frac{{\alpha {P_{total}}{{\left| {{h_{ab}}} \right|}^2}D_{jb}^\beta }}{{D_{jb}^\beta D_{ab}^\beta \sigma _b^2 + (1 - \alpha ){P_{total}}{{\left| {{h_{jb}}} \right|}^2}D_{ab}^\beta }})\} ,}\\
	{\begin{array}{*{20}{c}}
		{}
		\end{array}s.t\begin{array}{*{20}{c}}
		{}&{{C_1}:0 < \alpha  < 1,}
		\end{array}}\\
	{\begin{array}{*{20}{c}}
		{\begin{array}{*{20}{c}}
			{}&{{\kern 1pt} {\kern 1pt} {\kern 1pt} {\kern 1pt} {\kern 1pt} {\kern 1pt} {\kern 1pt} {\kern 1pt} {\kern 1pt} {\kern 1pt} {\kern 1pt}}
			\end{array}{C_2}}
		\end{array}:\mathop {\min }\limits_V ({P_{FA}} + {P_{MD}}) \ge 1 - \varepsilon .}
	\end{array}
	\end{eqnarray}
	After formulating the constraint $C_2$ (See Appendix B),  \eqref{q3} can be rewritten as follows
	\begin{eqnarray}\label{q4}
	\begin{array}{*{20}{l}}
	{\mathop {\max }\limits_\alpha  \{ R = \log (1 + \frac{{\alpha {P_{total}}{{\left| {{h_{ab}}} \right|}^2}D_{jb}^\beta }}{{D_{jb}^\beta D_{ab}^\beta \sigma _b^2 + (1 - \alpha ){P_{total}}{{\left| {{h_{jb}}} \right|}^2}D_{ab}^\beta }})\} ,}\\
	{\begin{array}{*{20}{c}}
		{}
		\end{array}s.t\begin{array}{*{20}{c}}
		{}&{{C_1}:0 < \alpha  < 1,}
		\end{array}}\\
	{\begin{array}{*{20}{c}}
		{\begin{array}{*{20}{c}}
			{}&{\,\,\,\,\,\,\,\,}
			\end{array}{C_2}}
		\end{array}:\frac{{(1 - \alpha )D_{je}^\beta }}{{(1 - \alpha )D_{ae}^\beta  - \alpha D_{je}^\beta }}\ln (\frac{{ \alpha D_{je}^\beta }}{{(1-\alpha) D_{ae}^\beta }}) < ln\varepsilon .}
	\end{array}
	\end{eqnarray}
	Since the optimization problem (22) is a non-convex one, we rewrite the objective function as
	follows 
	\begin{eqnarray}
	\Sigma (\alpha ) - \Psi (\alpha ),
	\end{eqnarray}
	where
	\begin{eqnarray}
	\left\{ \begin{array}{l}
	\Sigma (\alpha ) = log(D_{jb}^\beta D_{ab}^\beta \sigma _b^2 + (1 - \alpha ){P_{total }}D_{ab}^\beta {\left| {{h_{jb}}} \right|^2}\\
	\,\,\,\,\,\,\,\,\,\,\,\,\,\,\,\,\,\,\,\,\,\,\,\,\,\, + \alpha{P_{\max }}D_{jb}^\beta {\left\| {{h_{ab}}} \right\|^2})\\
	\Psi (\alpha ) = \log (D_{jb}^\beta D_{ab}^\beta \sigma _b^2 + (1 - \alpha ){P_{total }}D_{ab}^\beta {\left| {{h_{jb}}} \right|^2})
	\end{array} \right.
	\end{eqnarray}
	To solve the optimization problem, we use the DC method. As such, we approximate $\Psi (\alpha )$ as follows [16]
	\begin{eqnarray}
	\begin{array}{l}
	\Psi (\alpha ) \simeq \tilde \Psi (\alpha ) = \Psi (\alpha (\eta  - 1)) + \\
	{\nabla ^T}\Psi (\alpha (\eta  - 1))(\alpha  - \alpha (\eta  - 1)),
	\end{array}
	\end{eqnarray}
	where $\nabla $ is the gradient operator and ${\nabla ^T}\Psi (\alpha (\eta  - 1))$ is calculated as follows
	\begin{eqnarray}
	{\nabla ^T}\Psi (\alpha (\eta  - 1)) = \frac{{ - {P_{total }}D_{ab}^\beta {{\left| {{h_{jb}}} \right|}^2}}}{{D_{jb}^\beta D_{ab}^\beta \sigma _b^2 + (1 - \alpha (\eta  - 1)){P_{total }}D_{ab}^\beta {{\left| {{h_{jb}}} \right|}^2}}}.
	\nonumber\\
	\end{eqnarray}
	Finally, \eqref{q2} can be rewritten as $\Sigma (\alpha ) - \tilde \Psi (\alpha )$ which is a concave function. $C_2$ can be concave by using the change of value $T = (1 - \alpha )D_{ae}^\beta  - \alpha D_{je}^\beta .$
	Therefore, we have 
	\begin{eqnarray}\label{pp}
	\begin{array}{l}
	\mathop {\max }\limits_\alpha  \,\,\,\,\,\,\,\Sigma (\alpha ) - \tilde \Psi (\alpha )\\
	s.t\,\,\,\,{C_1}:0 < \alpha  < 1,\\
	\,\,\,\,\,\,\,\,\,\,{C_2}:(1 - \alpha )D_{ae}^\beta \ln (\frac{{ \alpha D_{je}^\beta }}{{(1-\alpha) D_{ae}^\beta }}) < Tln\varepsilon ,\\
	\,\,\,\,\,\,\,\,\,\,{C_3}:(1 - \alpha )D_{ae}^\beta  - \alpha D_{je}^\beta  < T.
	\end{array}
	\end{eqnarray}
	Since the optimization problem in \eqref{pp} is convex. we can use the available softwares such as CVX solver to solve it.
\end{subsection}
\begin{subsection}{Friendly Jamming Scenario}
	In this scenario, it is assumed that Bob can cancel the jamming signal. Hence, we have the following optimization problem 
	\begin{eqnarray}\label{lll}
	\begin{array}{l}
	\mathop {\max }\limits_\alpha  \{ R = log(1 + \frac{{\alpha {P_{total}}{{\left| {{h_{ab}}} \right|}^2}}}{{D_{ab}^\beta \sigma _b^2}})\} ,\\
	\,\,\,s.t\,\,\,\,\,\,\,\,\,\,{C_1},{C_2}.
	\end{array}
	\end{eqnarray}
	In \eqref{lll}, the objective function is convex. Therefore, we can use the CVX solver to solve it. In addition, we can  convert constraints $C_1$ and $C_2$ to convex constraints as in the previous scenario.
\end{subsection}
\section{Simulation Result}
In this section, our aim is to evaluate the secrecy performance of the proposed secure transmission scheme. For simplicity and without loss of generality, we assume that Alice, Bob, the jammer and the passive Eve are located at the positions (-2.5,2.5), (2.5,2.5), (2.5,-2.5) and (-2.5,-2.5), respectively. Unless otherwise stated, the network parameters are: number of antennas at Alice $ M_T=10$, number of antenna at Alice that are selected $N_D=6$, $\varepsilon=0.1$, $1-\varepsilon$ is lower bound of error detection probability, and the path-loss exponent $\beta=2$.

\begin{figure}[h!]
	\begin{center}
		\includegraphics[width=3.8in,height=2.9in]{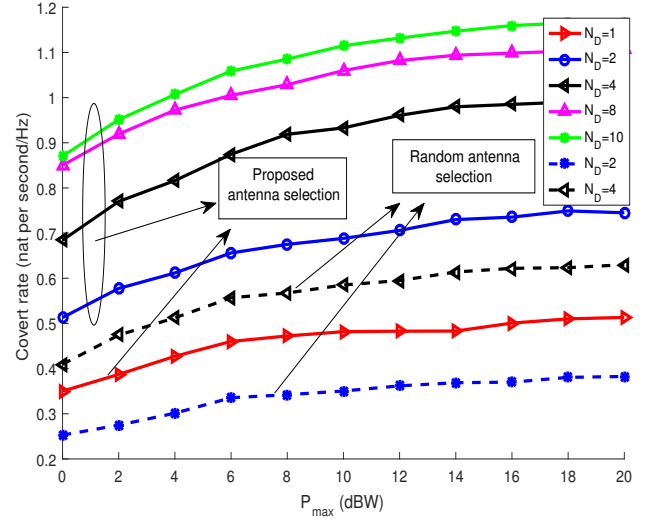}
		\caption{Covert rate versus the total transmit power.}
		\label{f1}
	\end{center}
\end{figure}

Fig. \ref{f1} shows the effect of the transmit power sent by Alice on the covert rate. According to simulation results, if the  transmit power increases, the covert rate increases. However, as can be seen, the covert rate ceiling is appeared at high transmit powers. This is because when the transmit power increases, jammer transmits jamming signal with higher power, hence, Alice can transmit information signal with higher power which leads to increase covert rate. Furthermore, we observe that by increasing the number of the best selected antennas for signal transmission, the achievable covert rate improves.  For example, if the transmit power is constant and the number of antennas which are selected for sending information signal is doubled, the covert rate is reduced $27\% $. However, the simulation results shows that the associated number of antennas should not be larger than 6 in order to balance the performance gain and implementation cost. If the number of antennas exceeds 6, in addition to Bob's condition, the Eve conditions will also improve for detection, which will make our hidden rate not change significantly. Since allocating large number of antennas for signal transmission imposes more hardware equipments (RF chains), higher latency and more much overhead to the network. Moreover, this figure compares random  antenna selection and proposed antenna selection.  It is necessary to note that random antenna selection with $N_D=4$ has more efficiency with respect to proposed method with $N_D=1$ and less efficiency with respect to proposed method with $N_D=2$. The reason is when the number of antennas increase diversity gain increases. But the proposed method with less diversity gain has more efficiency with respect to random antenna selection with higher diversity gain. The Fig. \ref{f1} shows proposed method enhances the covert rate approximately 2 times.
\begin{figure}[h!]
	\begin{center}
		\includegraphics[width=3.8in,height=2.9in]{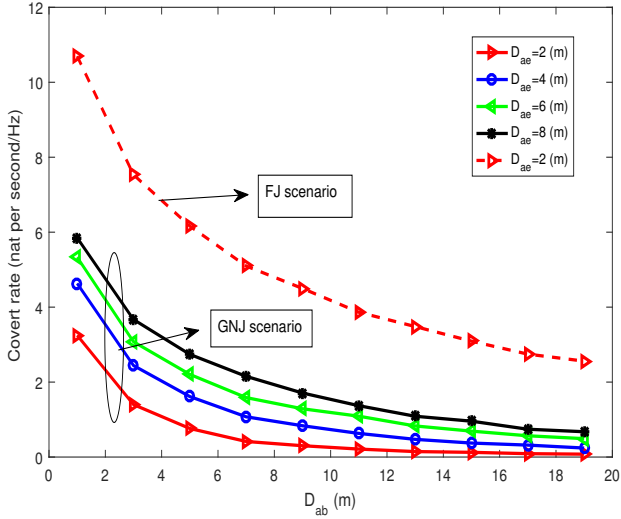}
		\caption{Covert rate versus the distance between Alice and Bob. We set $P_{total}=5W$.}
		\label{f3}
	\end{center}
\end{figure}

As we can see in Fig. \ref{f3}, the lower the distance between Alice and Bob, the higher the covert rate. In addition, the lower the distance between Alice and Eve, the lower the covert rate. Given Fig. \ref{f3}, if the distance between Alice and Bob  is increased by 6 meters, the covert rate reduces about $62\%$. Now, if the distance between Alice and Eve increases the same amount, the cover rate increases about 4.7 times. Thus, the impact of the distance between Alice and Eve is more important than the distance between Alice and Bob. Also, as shown in Figure 3, when the Bob can cancel the jamming signal, the covert rate increases dramatically. For example, if the distance between Alice and Jammer is 2 meters and Bob cancels the jamming signal, the covert rate is 5.6 times more.

\begin{figure}[h!]
	\begin{center}
		\includegraphics[width=3.7in,height=2.8in]{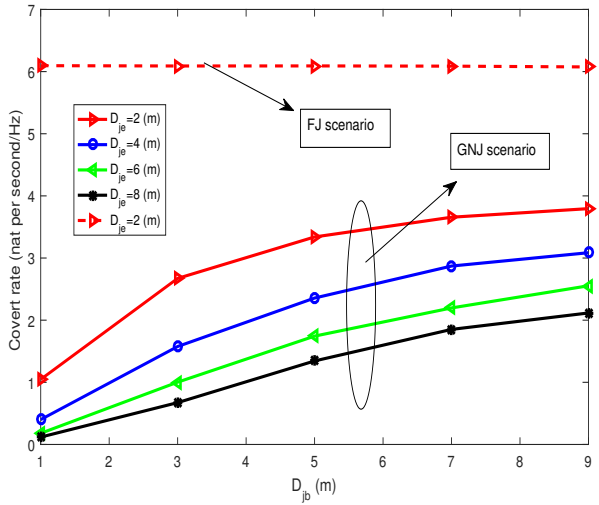}
		\caption{Covert rate versus the distance between the jammer and Bob. We set $P_{total}=5W$.}
		\label{f2}
	\end{center}
\end{figure}

Also, in Fig. \ref{f2}, we can see the effect of the distance between the jammer and Bob and Eve on the covert rate. According to simulation results, when Bob can not cancel the jamming signal, the lower the distance between the jammer and Bob, the lower the covert rate. For example, if the the distance between the jammer and Bob from 3 to 9 meters, the covert rate will increase $40\%$. Since the jamming signal treats as interference in Bob, by reducing the distance between the jammer and Bob, the covert rate is reduced. Moreover, the distance between the jammer and Eve is effective. If their distance is 6 meter high, the covert rate will decrease $63\%$. Also, when the jamming signal is canceled, increasing of the distance between Bob and jammer does not change the covert rate, and this parameter does not affect on the covert rate.

\section{Conclusion}
In the covert communication, the level of transmit power is low which lead to low SINR hence we aim to exploit from method without increasing the transmit power increase the covert rate. Our strategy to increase the covert rate is choosing the appropriate antenna in the transmitter to send the information. For this purpose, we proposed a system with a multi-antenna transmitter, a single-antenna jammer , a single-antenna receiver, and a single antenna eavesdropper. In this scenario, we encountered the optimization problem, which we use to solve the method of DC. Moreover,  we investigated two different scenarios: 1) GNJ where Bob is unable to cancel the jamming signal, 2) FJ scenario where Bob is able to cancel the jamming signal. As the simulation results show, by increasing the number of antennas which are selected in the transmitter and choosing an appropriate antenna for sending the information, it causes an increase in the covert rate. Furthermore, the distance between transmitter and the eavesdropper has more impact on the cover rate compared with the distance between the transmitter and the receiver.

\appendix
\hspace{3.15cm} Appendix A\\
It is simple to show that  ${{P_{FA}} + {P_{MD}}}$ is convex. In order to calculate the minimum value of ${{P_{FA}} + {P_{MD}}}$, we derivative  ${{P_{FA}} + {P_{MD}}}$ with respect to V and then we find the root of it. Following this, we obtain the optimal $V^*$ as follows
\begin{eqnarray}
\begin{array}{l}
{V^*} = \frac{{{\phi _0}{\phi _1}}}{{{\phi _1} - {\phi _0}}}Ln\frac{{{\phi _1}}}{{{\phi _0}}} + \sigma _e^2.
\end{array}
\end{eqnarray}
\hspace{3.3cm}  Appendix B\\
For the constraint $C_2$, we can write
\begin{eqnarray}
\begin{array}{l}
\frac{{\alpha D_{je}^\beta }}{{(1 - \alpha )D_{ae}^\beta  - \alpha D_{je}^\beta }}\begin{array}{*{20}{c}}
\times 
\end{array}\left[ {e\frac{{(1 - \alpha )D_{je}^\beta }}{{(1 - \alpha )D_{ae}^\beta  - \alpha D_{je}^\beta }}ln(\frac{{\alpha D_{je}^\beta }}{{(1 - \alpha )D_{ae}^\beta }})} \right.\\
\left. {\,\,\,\,\,\,\,\,\,\,\,\,\,\,\,\,\,\,\,\,\,\,\,\,\,\,\,\,\,\,\,\,\,\,\,\,\,\,\,\,\,\,\,\,\,\, - e\frac{{\alpha D_{je}^\beta }}{{(1 - \alpha )D_{ae}^\beta  - \alpha D_{je}^\beta }}ln(\frac{{\alpha D_{je}^\beta }}{{(1 - \alpha )D_{ae}^\beta }})} \right] 
\end{array}
\nonumber
\end{eqnarray}
\begin{align}\label{ppo}
\begin{array}{l}
= \frac{{\alpha D_{je}^\beta }}{{(1 - \alpha )D_{ae}^\beta  - \alpha D_{je}^\beta }} \times {e^{{{\frac{{(1 - \alpha )D_{ae}^\beta }}{{(1 - \alpha )D_{ae}^\beta  - \alpha D_{je}^\beta }}}^{ln(\frac{{\alpha D_{je}^\beta }}{{(1 - \alpha )D_{ae}^\beta }})}}}}\\
\,\,\,\,\,\,\,\,\, \times \,\,\left. {\left[ {1 - {e^{ln(\frac{{\alpha D_{je}^\beta }}{{(1 - \alpha )D_{ae}^\beta }})}}} \right.} \right] >  - \varepsilon .
\end{array}
\end{align}
Therefore, we can rewrite the inequality \eqref{ppo} as
\begin{eqnarray}
{{e^{\frac{{(1 - \alpha )D_{ae}^\beta }}{{(1 - \alpha )D_{ae}^\beta  - \alpha D_{je}^\beta }}}}^{ln(\frac{{\alpha D_{je}^\beta }}{{(1 - \alpha )D_{ae}^\beta }})} < \varepsilon .}
\nonumber
\end{eqnarray}
which is equivalent to the constraint $C_2$.

\end{document}